\documentclass[aps,twocolumn,prl,letterpaper,superscriptaddress,showpacs,nofootinbib]{revtex4}
\usepackage{amsmath,amssymb}
\usepackage{color}
\usepackage[pdftex]{graphicx} 
\usepackage{hyperref}

\begin{document}


\newcommand{\be}{\begin{equation}}
\newcommand{\ee}{\end{equation}}
\newcommand{\R}[1]{\textcolor{red}{#1}}
\newcommand{\B}[1]{\textcolor{blue}{#1}}


\title{Narrowing the filter cavity bandwidth via optomechanical interaction}

\author{Yiqiu Ma}
\email{myqphy@gmail.com}
\affiliation{School of Physics, University of Western Australia, WA 6009, Australia}
\author{Shtefan L. Danilishin}
\affiliation{School of Physics, University of Western Australia, WA 6009, Australia}
\author{Chunnong Zhao}
\email{chunnong.zhao@uwa.edu.au}
\affiliation{School of Physics, University of Western Australia, WA 6009, Australia}
\author{Haixing Miao}
\email{haixing@caltech.edu}
\affiliation{Theoretical Astrophysics 350-17, California Institute of Technology, Pasadena,
CA 91125, USA}
\author{W. Zach Korth}
\affiliation{LIGO Laboratory, California Institute of Technology, Pasadena, California
91125 USA}
\author{Yanbei Chen}
\affiliation{Theoretical Astrophysics 350-17, California Institute of Technology, Pasadena,
CA 91125, USA}
\author{Robert L. Ward}
\affiliation{Centre for Gravitational Physics, Department of Quantum Science, Australian National University, Canberra ACT 0200 Australia}
\author{D. G. Blair}
\affiliation{School of Physics, University of Western Australia, WA 6009, Australia}

\begin{abstract}
We propose using optomechanical interaction to narrow the bandwidth of filter cavities for
achieving frequency-dependent squeezing in advanced gravitational-wave detectors, inspired
by the idea of optomechanically induced transparency. This can allow us to achieve a cavity bandwidth on the order of one hundred Hz using small-scale cavities. Additionally, in contrast to a passive Fabry-P\'{e}rot cavity, the resulting cavity bandwidth can  be dynamically tuned, which is useful for adaptively optimizing the detector sensitivity when switching amongst different operational modes. The experimental challenge for its implementation is
a stringent requirement for very low thermal noise of the mechanical oscillator, which would need superb mechanical quality factor and very low temperature. We consider one possible setup to relieve this requirement by using optical dilution to enhance the mechanical quality factor.
\end{abstract}

\maketitle


{\it Introduction.---}Advanced interferometric gravitational wave (GW) detectors, e.g., the advanced LIGO\,\cite{Abbott2009}, advanced VIRGO\,\cite{aVir2009} and KAGRA\,\cite{Somiya2012}, are expected to be limited by quantum noise over almost the entire detection band. Further enhancement of the detector sensitivity requires manipulation of the optical field and the readout scheme at the quantum level. One approach proposed by Kimble {\it et al.}\,\cite{Kimble2001} is injecting frequency-dependent squeezed light into the main interferometer. A series of optical cavities is used to filter the squeezed light and to create proper rotation of the squeezing angle at different frequencies. In order to achieve a broadband reduction of quantum noise, the frequency scale of these filter cavities needs to match that of quantum noise of the main interferometer. For the advanced LIGO, the quantum noise is dominated by quantum radiation pressure noise at low frequencies and shot noise at high frequencies---the transition happens around 100Hz, which determines the required filter cavity bandwidth.

The original proposal in Ref.\,\cite{Kimble2001} is using filter cavities of kilometer length. Recently, Evans {\it et.al} \,\cite{Evans2013} proposed a more compact (10 meters) filter cavity with $10^5$ finesse to achieve the required bandwidth. With such a high finesse, a small optical losses can degrade the squeezing and become the key limiting factor in the filter cavity performance.  Isogai {\it et.al} have experimentally demonstrated that the optical losses from current mirror technology are sufficiently small to build such a filter cavity that will be useful for the advanced LIGO \,\cite{Isogai2013}. However if we want to further increase the compactness of the filter cavity, then the requirement for optical loss becomes more stringent. In this case, one solution is to go beyond the paradigm of passive cavities. One proposed approach is to actively narrow the cavity bandwidth by using  electromagnetically induced transparency (EIT) effect in a pumped atomic system\,\cite{Mikhailov2006}. In principle, the cavity can be made to be on the centimeter-scale while still having a bandwidth comparable to a much longer high-finesse cavity.  Additionally, with an active element, the cavity optical properties can be dynamically tuned by changing the power of the control pumping field. This has the advantage of allowing optimization of the filter cavity for different operational modes of the detector, where the quantum noise has different frequency dependencies, e.g., tuned vs. detuned resonant sideband extraction (RSE) in the case of the advanced LIGO.

\begin{figure}[!t]
\includegraphics[width=0.33\textwidth]{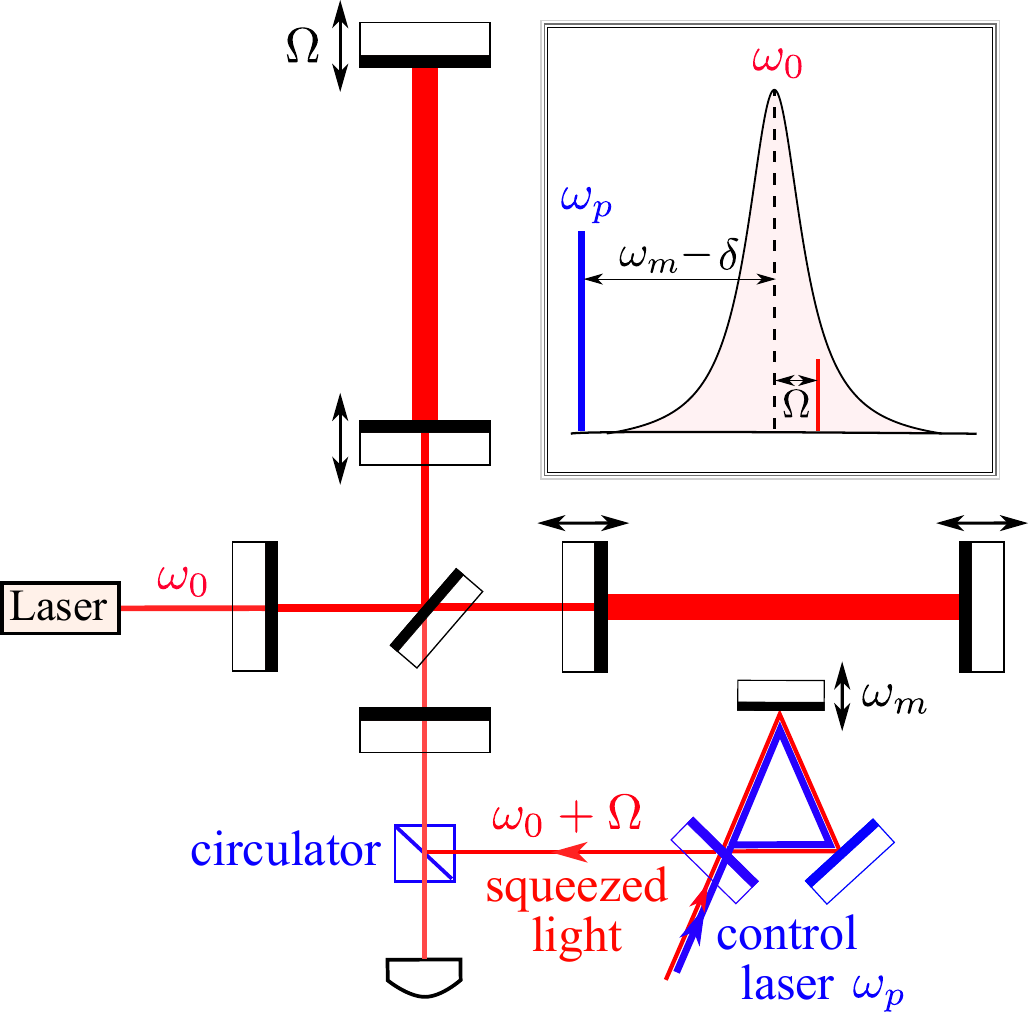}
\caption{(Color online) Schematic showing the configuration for achieving frequency-dependent
squeezing.
\label{fig_config}}
\end{figure}

The active atomic system is generally lossy, which will degrade the squeezing level. Here we propose to narrow the filter cavity bandwidth using the optomechanical analogue of EIT,  optomechanically induced transparency (OMIT), which has recently been experimentally demonstrated by Weis {\it et al.}\,\cite{Weis2010} and Teufel {\it et al.}\,\cite{Teufel2011}. In comparison with these OMIT experiments, we consider a different parameter regime and use an overcoupled cavity to attain the desired performance. The scheme integrated with the main interferometer is illustrated in Fig.\,\ref{fig_config}. The filter cavity consists of a mirror-endowed
mechanical oscillator with eigenfrequency $\omega_m$ that is much larger than the cavity bandwidth $\gamma$.
A control pump laser drives the filter cavity at frequency $\omega_p$, detuned from the cavity
resonant frequency $\omega_0$ (also the laser frequency of the main interferometer)
by $\omega_m-\delta$ with $\delta$ comparable to the gravitational-wave signal frequency $\Omega$. As we will show, the optomechanical interaction modifies the cavity
response and gives rise to the following input-output relation for the sideband
at $\omega_0+\Omega$:
\be\label{eq_io}
\hat a_{\rm out}(\Omega)\approx \frac{\Omega-\delta-i\gamma_{\rm opt}}
{\Omega-\delta+i\gamma_{\rm opt}}\hat a_{\rm in}(\Omega)+\hat n_{\rm th}(\Omega)\,,
\ee
where $\gamma_{\rm opt}$ is defined as:
\be
\gamma_{\rm opt}= \frac{4P_c\omega_0}{m\omega_m c^2T_f}\,,
\ee
with $P_c$ being the intra-cavity power of the control field, $m$ the mass of the mechanical oscillator and $T_f$ the transmissivity of the front mirror (the end mirror is totally reflective). The first term in Eq.\ref{eq_io} gives the input-output relation of a standard optical cavity with the original cavity bandwidth $\gamma$
 replaced by $\gamma_{\rm opt}$, which can be significantly smaller than $\gamma$ as well as dynamically tuned by changing the control beam power.

The second term $\hat n_{\rm th}$ arises from the thermal fluctuation of the mechanical oscillator. It
is uncorrelated with the input optical field $\hat a_{\rm in}$ and therefore decohering the squeezed light. In order for its effect to be small, we require:
\be\label{eq_cond}
\frac{8k_B T}{Q_m} < \hbar \gamma_{\rm opt}\,,
\ee
with $Q_m$ the mechanical quality factor and $T$ the environmental temperature. Given
the fact that the desired effective cavity bandwidth is
$\gamma_{\rm opt}/2\pi\approx 100$ Hz, we have
\be\label{eq_cond2}
\frac{T}{Q_m}<6.0\times 10^{-10}\,{\rm K}\,.
\ee
This is challenging to achieve even with low-loss materials at cryogenic temperature.
A possible solution is to use optical dilution, first proposed by Corbitt {\it et al.}\,\cite{Corbitt2007, Corbitt2007b,Chang2012,Ni2012}. It allows for a significant boost of the mechanical quality factor by using the optical field, to provide most of the restoring force. Later we illustrate its applicability for our purpose.

{\it Optomechanical dynamics.---}Here we provide the details behind Eq.\,\eqref{eq_io} by
analyzing the dynamics of the optomechanical filter cavity, starting from the standard linearized Hamiltonian\,\cite{Wilson-Rae2007,Marquardt2007}:
\begin{align}\nonumber
\hat {\cal H}=&\hbar\,\omega_0 \hat a^{\dag}\hat a+\frac{\hat p^2}{2m}+\frac{1}{2}m\,\omega_m^2\hat x^2+\hbar\,\bar G_0
\hat x(\hat a^{\dag}+\hat a)\\&+i\hbar\sqrt{2\gamma}(\hat a^{\dag}\hat a_{\rm in}e^{-i\omega_p t}-\hat a\,\hat a_{\rm in}^{\dag}e^{i\omega_p t})\,.
\end{align}
In the Hamiltonian, $\hat a$ is the annihilation operator of the cavity mode and $\hat a_{\rm in}$ is the annihilation operator for the input optical field (the squeezed light in our case); $\hat x$ $(\hat p)$ is the oscillator position (momentum); $\bar G_0=[2P_c \omega_0/(\hbar c L)]^{1/2}$ with $L$ being the cavity length.
In the rotating frame at frequency $\omega_p$ of the control laser, the Heisenberg equation of motion reads:
\begin{align}
m(\ddot{\hat x}+\gamma_m \dot {\hat x}+\omega_m^2 \hat x) &=-\hbar\,\bar G_0(\hat a+\hat a^{\dag})+\hat F_{\rm th},\\\label{eq_aomega}
\dot{\hat a}+(\gamma+i\Delta)\hat a &=-i\,\bar G_0 \hat x+\sqrt{2\gamma}\,\hat a_{\rm in},
\end{align}
where $\Delta\equiv \omega_0-\omega_p$ is the detuning frequency and we have included the mechanical damping and associated Langevin force $\hat F_{\rm th}$.
Solving these equations of motion in the frequency domain yields
\begin{align}\label{eq_x}
\hat x(\omega)&=\chi_m(\omega)\left\{\hbar \bar G_0 [\hat a(\omega)+\hat a^{\dag}(-\omega)]+\hat F_{\rm th}(\omega)\right\},\\\label{eq_a}
\hat a(\omega)&=\chi_c(\omega)[-i\bar G_0\,\hat x(\omega)+\sqrt{2\gamma}\,\hat a_{\rm in}(\omega)]\,.
\end{align}
We have defined the susceptibilities $\chi_m\equiv-[m(\omega^2-\omega_m^2+i\gamma_m\omega)]^{-1}$ and $\chi_c\equiv [\gamma-i(\omega-\Delta)]^{-1}$.

{\it Relevant parameter regime.---}We consider the parameter regime leading to Eq.\,\eqref{eq_io}. This requires $\Delta = \omega_m -\delta$ with $\omega_m
\gg \delta$, and the so-called resolved-sideband regime $\omega_m\gg\gamma$.
Correspondingly, the lower sideband of the cavity mode $\hat a(-\omega)$ in Eq.\,\eqref{eq_x}
is negligibly small and can be ignored (we will analyze the effect of this approximation later).
We therefore obtain [cf. Eqs.\,\eqref{eq_x} and \eqref{eq_a}]:
\be
\hat a(\omega)\approx \frac{\sqrt{2\gamma}\,\hat a_{\rm in}(\omega) -i\bar G_0\chi_m(\omega)\hat F_{\rm th}(\omega) }{ \chi^{-1}_c(\omega)+i\,\hbar \,\bar G_0^2\,\chi_m(\omega)}\,.
\ee
Since we are interested in the signal sidebands around
$\omega_0$, we rewrite the above
expression in terms of $\Omega$
by using the equality $\omega=\Delta + \Omega$ [cf. the inset of Fig.\,\ref{fig_config}]. Given $\Omega\approx \delta \ll\omega_m$, we have $\chi_m \approx -[2m\omega_m(\Omega-\delta+i\gamma_m)]^{-1}$ and $\chi_c \approx \gamma^{-1}\,. $
Together with $\hat a_{\rm out}=-\hat a_{\rm in}+\sqrt{2\gamma}\, \hat a$, we obtain
\be
\hat a_{\rm out}(\Omega)\approx \frac{\Omega-\delta+i\gamma_m - i\gamma_{\rm opt}}{\Omega-\delta+i\gamma_m + i\gamma_{\rm opt}}\hat a_{\rm in}(\Omega)+\hat n_{\rm th}(\Omega)
\ee
with the additional noise term $\hat n_{\rm th}$ defined as
\be\label{eq_nth}
\hat n_{\rm th}(\Omega)=\frac{i\sqrt{2\gamma}\,\gamma_{\rm opt}\hat F_{\rm th}(\Omega)}{\hbar \bar G_0(\Omega-\delta+i\gamma_m - i\gamma_{\rm opt})}.
\ee
For a high quality factor oscillator $\gamma_m\ll \gamma_{\rm opt}$, we can ignore
$\gamma_m$ and recover the input-output relation shown in Eq.\,\eqref{eq_io}.

To maintain coherence of the squeezed light, the fluctuations due to the thermal noise term $\hat n_{\rm th}$ need to be much smaller than those due to the input field; equivalently, the quantum radiation pressure noise on the mechanical oscillator
from the squeezed light needs to dominate over thermal noise of the oscillator. Given the fact that
$\langle \hat F_{\rm th}^{\dag}(\Omega)\hat F_{\rm th}(\Omega')\rangle=4m\gamma_m k_B T\delta(\Omega-\Omega')$, the requirement on the noise spectrum for $\hat n_{\rm th}$ reads
\be
\label{eq_Sth}
S_{\rm th}(\Omega)=\left(\frac{8k_B T}{\hbar \gamma_{\rm opt}Q_m}\right)\frac{\gamma_{\rm opt}^2}
{(\Omega-\delta)^2+\gamma_{\rm opt}^2}<1\,.
\ee
The thermal noise effect is maximal around $\Omega\sim \delta$, from which
we obtain the condition shown in Eq.\,\eqref{eq_cond}.

{\it Effects of optical loss and finite cavity bandwidth.---}Apart from the above-mentioned
thermal noise, there are other decoherence effects: (i)
the additional radiation pressure noise introduced by the optical loss, and also (ii) the effect of the lower sideband due to the finite cavity bandwidth, ignored in the resolved-sideband limit. Their effects are similar to the above thermal force noise; therefore we can quantify their magnitude using the noise spectrum referred to the output. For the optical loss,
\be
S_{\epsilon}(\Omega)=\left(\frac{c \,\epsilon}{\gamma L}\right)\frac{\gamma_{\rm opt}^2}{(\Omega-\delta)^2+\gamma_{\rm opt}^2}\,,
\ee
where $\epsilon$ is the magnitude
of the optical loss (e.g., $\epsilon=10^{-5}$ for 10ppm loss).
Similarly, for the contribution from the lower sideband, we have
\be
S_{-\omega_m}(\Omega)= \left(\frac{\gamma}{\omega_m}\right)^2\frac{\gamma_{\rm opt}^2}
{(\Omega-\delta)^2+\gamma_{\rm opt}^2}\,.
\ee
These two need to be taken into account when estimating the performance of this optomechanical filter cavity.


\begin{figure}[!t]
\includegraphics[width=0.35\textwidth]{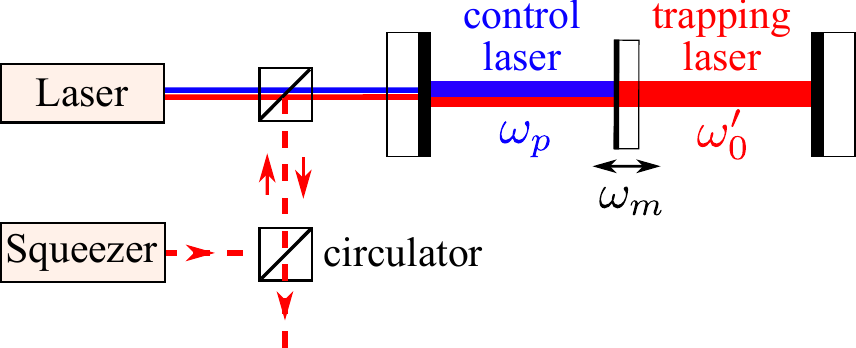}
\caption{Schematics of the coupled cavity setup for the filter cavity.
\label{fig_scheme}}
\end{figure}

{\it Possible experimental scheme.---}We have shown in Eq.\,\eqref{eq_cond2} that the most significant issue is
the thermal noise, which puts a stringent requirement
on the mechanical system and the environmental temperature. As we mentioned earlier, One possible way to mitigate this is
using optical dilution explored by Corbitt {\it et al.}, in which the optical restoring force is due to  the linear dependence of radiation pressure force
on the oscillator position. This scheme has a limitation from quantum back action noise associated with a linear position response. Korth {\it et al.}\,\cite{Korth2012}, have showed how measurement-based feedback can cancel the quantum back action. Such a cancellation is, however, limited by quantum efficiency of the photodiode for measurement.

Here we consider optical dilution using a coupled cavity scheme, shown in Fig.\ref{fig_scheme}, with a mirror-endowed oscillator placed in the middle of a Fabry-P\'erot cavity, first implemented by Thompson {\it et al.}\,\cite{Thompson2008, Jayich2008}. Interestingly, this scheme allows for an internal cancellation of the quantum back action associated with a linear optical spring, and thus it avoids the limitation of the scheme in Ref.\,\cite{Korth2012}. A detailed analysis is given in the supplemental material\,\cite{Supplemental}. An intuitive picture behind this back-action evasion effect can be described as follows. The optical field on the left-hand side of the middle oscillator consists of two parts: (i) the immediate reflection from the oscillator and (ii) the transmitted field from the right-hand side, both containing the position information of the oscillator. The coupled cavity has a doublet resonance. It turns out that, when the trapping field is resonantly tuned to one of the doublet and the end mirror is perfectly reflective, the position information from these two parts destructively interfere, resulting in a cancelation of the back action.

Strong trapping beam can induce an optical spring frequency $\omega_{\text{opt}}\gg\omega_{m0}$ with:
\begin{equation}\label{eq_opt}
\omega_{\text{opt}}^2=\frac{2P_{\rm trap}\omega_0'}{m c^2 \sqrt{T_s} T_f}\,,
\end{equation}
where $P_{\text{trap}}$ and  $\omega_0'$ are the input power and optical frequency of the trapping beam, the $T_s$ and $T_f$ are the transmissivity of the mirror-endowed oscillator and the front mirror, respectively. The modified quality factor can be greatly boosted since the mechanical dissipation rate $\gamma_m$ is unchanged.

This optical dilution scheme also has its own limitations. Firstly, in reality there is no perfectly reflective mirror and always some optical loss, and so the above-mentioned cancellation cannot be perfect.  The residual radiation pressure noise, referred to the output, is given by:
\begin{equation}\label{eq_rpnoise}
S_{\epsilon}^{\rm opt}(\Omega)= \frac{4\omega_0' P_{\rm trap}\epsilon}{m \gamma_{\rm opt}\omega_m c^2 T_s T_f}\frac{\gamma_{\rm opt}^2}{(\Omega-\delta)^2+\gamma_{\rm opt}^2}\,.
\end{equation}
Secondly, the optical spring effect is frequency dependent: $K_{\rm opt}(\omega)\approx m\omega_{\rm opt}^2-i m \Gamma\omega-m_{\rm opt}\omega^2$. This tells us that the optical spring can modify not only the resonance frequency, but also the mechanical damping and the effective inertia (mass), which could induce instability. Lastly, finite absorption of the laser power in the oscillator will increase its temperature and may increase the thermal noise. The size of this effect, however, depends on the mechanical structure and the detailed loss mechanism.

\begin{figure}[!b]
\includegraphics[width=0.4\textwidth]{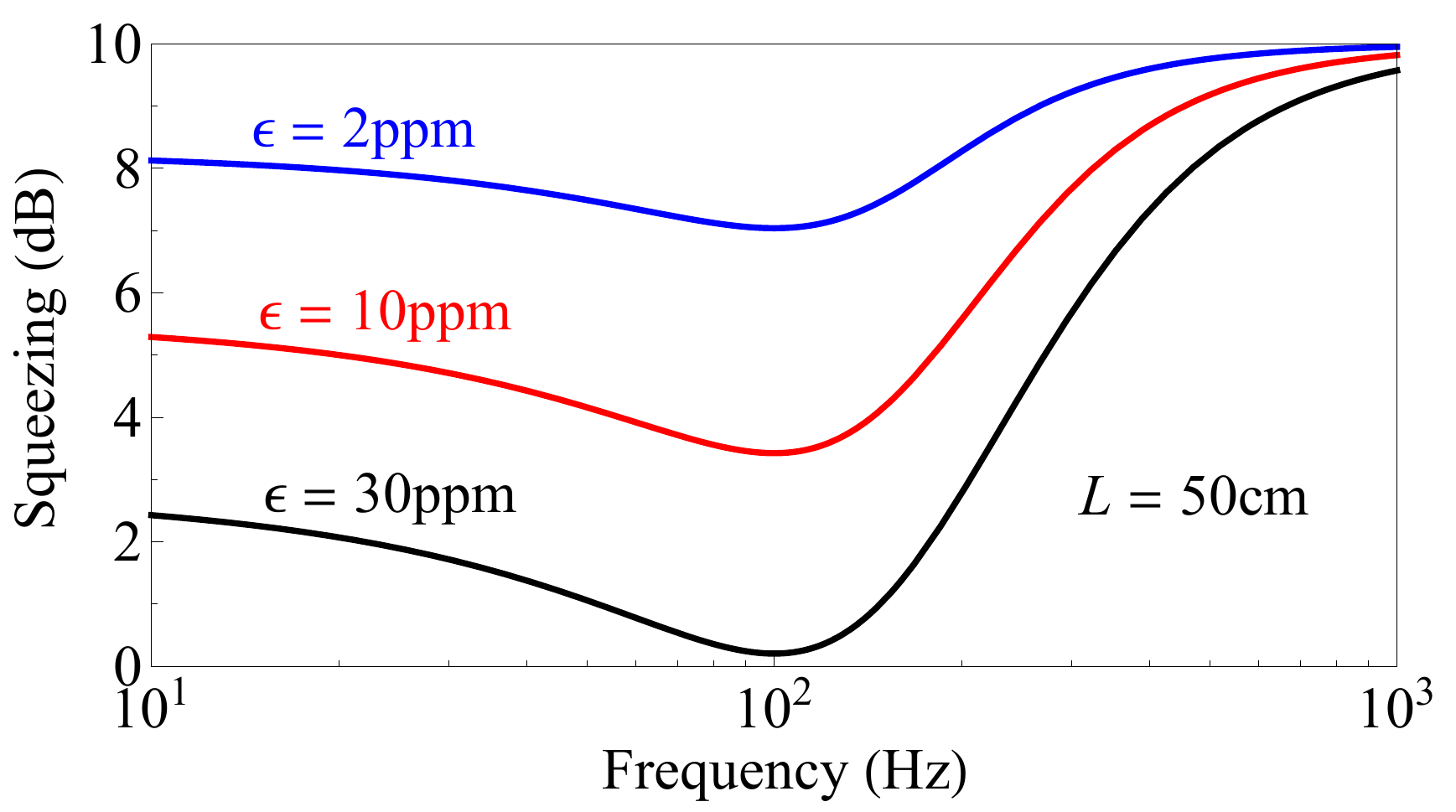}
\caption{Resultant squeezing level from injecting 10 dB of input squeezing into an optomechanical filter cavity using the optical-dilution scheme in Fig.\,\ref{fig_scheme}, with parameters in Table\,\ref{table:sample1}, for several values of the optical loss $\epsilon$.
\label{fig_sqz_deg}}
\end{figure}

{\it An example.---}We illustrate the requirements for experimentally realizing the
optomechanical filter cavity using optical dilution shown in Fig.\,\ref{fig_scheme} with some example parameters in Table\,\ref{table:sample1}. These values are chosen after considering the above mentioned effects, which can cause decoherence to the squeezed light, such that
\begin{equation}\label{eq:final_cond}
S^{\rm max}_{\rm tot}=S_{\rm th}(\delta)+S_{\epsilon}(\delta)+S_{-\omega_m}(\delta)+S_{\epsilon}^{\rm opt}(\delta)<1\,.
\end{equation}
In addition, once we fix the oscillator mass $m$ and transmissivity $T_s$, we can minimize $S^{\rm max}_{\rm tot}$ by looking into the scaling of different parameters, which determines the trapping beam power $P_{\rm trap}$, the front mirror transmissivity $T_f$, and the environmental temperature $T$. We end up with the following scaling of $S^{\rm max}_{\rm tot}$ in terms of optical loss and cavity length:
\begin{equation}\label{eq:scaling}
 S^{\rm max}_{\rm tot}\approx 3\times10^3 \epsilon^{4/5}/L^{2/5}\,.
\end{equation}
The resulting degradation to the squeezing factor due to optical loss is shown in Fig.\,\ref{fig_sqz_deg} for a cavity length of 50cm. In comparison to a passive filter cavity for which the performance degrades as $\epsilon/L$\,\cite{Isogai2013}, the optomechanical filter cavity using the optical-dilution scheme has a milder dependence on $L$, which yields the possibility of being small scale.

\begin{table}[!t]
\caption{Example parameter values}
\begin{tabular}{ccc}
\hline
\hline
Parameter & Description & Value\\
\hline
$L$ & filter cavity length &50{\rm cm} \\
\hline
$T_f$& front mirror transmissivity &250ppm\\
\hline
$T_s$&transmissivity of oscillator & 3000ppm\footnote{This value is only for the trapping beam; for the control field, the value is close to unity (limited by optical loss), requiring a dichroic coating.}\\
\hline
$P_{\text{trap}}$&trapping beam input power&1.6mW\\
\hline
$\lambda_0'$ & trapping beam wavelength & 532nm \\
\hline
$m$&oscillator mass &500ng\\
\hline
$\omega_{m_0}/(2\pi)$&bare mechanical frequency & 200Hz\\
\hline
$Q_{m_0}$& bare mechanical quality factor &$ 10^{8}$\footnote{According to~\cite{Cagnoli2000}, the mechanical damping of
some material structures are as small as $10^{-6}$Hz, which sets this possible value.}\\
\hline
$T$&environmental temperature&1K\\
\hline
$P_c$&control beam intra-cavity power&0.1mW\\
\hline
$\lambda_0$ & control beam wavelength & 1064nm\\
\hline
$\gamma_{\rm opt}/(2\pi)$ & effective cavity bandwidth & 100Hz\\
\hline
\hline
\end{tabular}
\label{table:sample1}
\end{table}

\begin{table}[!t]
\caption{Effective oscillator parameters}
\begin{tabular}{ccc}
\hline
\hline
$\omega_{\rm opt}/(2\pi)$& optical spring frequency [Eq.\eqref{eq_opt}] &20kHz\\
\hline
$Q_m$& final mechanical quality factor & $2\times 10^{10}$\\
\hline
$\Gamma/(2\pi)$&optical (anti-)damping rate [Eq.(A.9)] & $-8$mHz\\
\hline
$m_{\rm opt}$& negative optical inertia [Eq.(A.10)] & $-8.5$pg\\
\hline
\hline
\end{tabular}
\label{table:sample2}
\end{table}

The mechanical dynamics are modified by the opto-mechanical interaction, and the new effective parameters of the oscillator are summarised in Table\,\ref{table:sample2}. The optical spring shifts the mechanical resonant frequency from its bare value of 100Hz to 20kHz, which results in a two hundred fold increase in the quality factor. Comparing Table\,\ref{table:sample1} and Table\,\ref{table:sample2}, we can see that the negative optical damping and inertia do not pose an important problem.

We would like to point out that this scheme might not function as expected due to heating from finite absorption of the laser power. The intra-cavity power of the trapping beam, given the listed parameter values, is around 10W.  For 10\,ppm absorption, this amounts to $0.1$mW of heat deposited into the nano-mechanical oscillator. We make an order-of-magnitude estimate in the supplemental material and find this can create a nonuniform temperature distribution with a maximum around 10K near the beam spot. Further detailed study is required to estimate how this nonuniform temperature distribution on the oscillator affects the total thermal noise. Specifically in this case, the dissipation mainly comes from the clamping point where the temperature is still low. If this nonuniform temperature distribution indeed introduces significant thermal noise, then alternative materials with higher thermal conductivity at low temperature would need to be manufactured.


{\it Conclusion.---}We have considered the use of optomechanical interaction to narrow the bandwidth of a filter cavity for frequency-dependent squeezing in future advanced gravitational-wave detectors. However, due to susceptibility to thermal decoherence, its feasibility is conditional on advancements in low-loss mechanics and optics.


{\it Acknowledgements.---}We thank Huan Yang, David McCelland, Farid Khalili, Li Ju and Jiayi Qin for
fruitful discussions. Y.M., S.D., C.Z., R.W., and D.B. have been supported by
the Western Australia Centers of Excellence program, and by the Australian
Research Council; W.Z.K. is supported by NSF Grant PHY-0757058;
H.M.\ and Y.C.\ are supported by NSF Grants PHY-1068881 and CAREER
Grant PHY-0956189.


In this supplemental material, we will show additional
details and derivations for the
optomechanical dynamics of the optical-dilution scheme shown in Fig.\,2 of the main text.

\renewcommand{\theequation}{A.\arabic{equation}}
\renewcommand{\thefigure}{A.\arabic{figure}}
\setcounter{equation}{0}
\subsection{Hamiltonian and equation of motion}
The Hamiltonian of the system can be written as:
\begin{equation}
\begin{split}
H=\hbar\omega_0(\hat{a}^{\dagger}\hat{a}+\hat{b}^{\dagger}\hat{b})+\frac{\hat{p}^2}{2m}+\frac{1}{2}m\omega_m^2\hat{x}^2+\hbar\omega_s(\hat{a}^{\dagger}\hat{b}+\hat{a}\hat{b}^{\dagger})\\+
\hbar G_0\hat{x}(\hat{a}^{\dagger}\hat{a}-\hat{b}^{\dagger}\hat{b})+H^{opt}_{ext}+H^{m}_{ext}.
\end{split}
\end{equation}
Here, $\hat{a},\hat{b}$ are annihilation operators for cavity modes in left and right sub-cavity (with resonant frequency $\omega_c$) respectively. $\hat{x},\hat{p}$ are the position and momentum operators of the vibrating mirror. $\omega_s$ is the coupling constant for $\hat{a}$ and $\hat{b}$ and $G_0$ is defined to be $\omega_0/L$. $H^{opt}_{ext}=i\hbar\sqrt{2\gamma_f}(\hat a^{\dagger}\hat a_{\text{in}}-\text{h.c})+i\hbar\sqrt{2\gamma_\epsilon}(\hat b^{\dagger}\hat b_{\text{in}}-\text{h.c})$ and $H^{m}_{ext}$ correspond to the coupling of the system to the environment.

\begin{figure}[!b]
\includegraphics[width=0.4\textwidth]{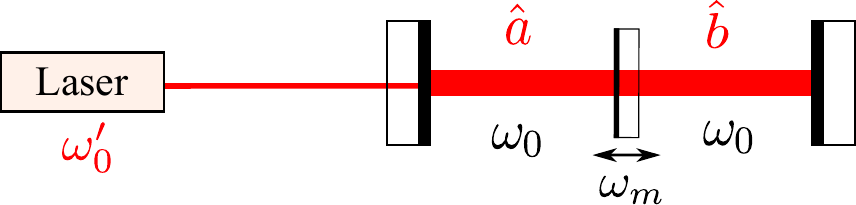}
\caption{Basic configuration of the proposed
scheme: a vibrating mirror trapped in a Fabry-P\'{e}rot cavity.
$\hat a$ and $\hat b$ are the light field operators in the left and
right subcavities, respectively.
\label{membraneinmiddle}}
\end{figure}

The Heisenberg equations of motion in the rotating frame of the trapping beam at frequency $\omega_0'$ can be derived as:
\begin{subequations}
\begin{align}
&\dot{\hat{a}}=i\Delta_t\hat{a}-\gamma_f\hat{a}-i\omega_s\hat{b}-iG_0\hat{x}\hat{a}+\sqrt{2\gamma_f}\hat{a}_{\text{in}},\\
&\dot{\hat{b}}=i\Delta_t\hat{b}-\gamma_\epsilon\hat{b}-i\omega_s\hat{a}+iG_0\hat{x}\hat{b}+\sqrt{2\gamma_\epsilon}\hat{b}_{\text{in}},\\
&\dot{\hat{p}}=-m\omega_m^2\hat{x}-\gamma_m\hat{p}-\hbar G_0(\hat{a}^{\dagger}\hat{a}-\hat{b}^{\dagger}\hat{b})+F_{\text{th}},\\
&\dot{\hat{x}}=\hat{p}/m.
\end{align}
\end{subequations}
Here, $\gamma_f=cT_f/4L$ and $\gamma_\epsilon=c\epsilon/4L$, $T_f$ and $\epsilon$ are the transmissivity of the front mirror and the loss of the system through the end mirror, $\Delta_t=\omega_0'-\omega_0$ is the detuning of the pumping laser field with respect to the half-cavity resonance. Suppose we pump the cavity by injecting a laser field through the front mirror (\emph{single-side pumping}), then $\bar{b}_{\text{in}}=0$. These equations can be solved perturbatively. The zeroth order terms give us the classical amplitude of the intra-cavity mode in both sub-cavities and the first order terms carry information about the mirror vibration along with quantum noise due to the non-zero transmissivity of the cavity end mirror.

From the above Heisenberg equations of motion, we have the steady state fields in
the two sub-cavities:
\begin{subequations}\label{classfield}
\begin{align}
\bar{a}=\frac{(i\Delta_t-\gamma_\epsilon)\sqrt{2\gamma_f}\bar{a}_{\text{in}}}
{\Delta_t^2-\omega_s^2-\gamma_f\gamma_\epsilon+i\Delta_t(\gamma_f+\gamma_\epsilon)}\\
\bar{b}=\frac{i\sqrt{2\gamma_f}\omega_s\bar{a}_{\text{in}}}
{\Delta_t^2-\omega_s^2-\gamma_f\gamma_\epsilon+i\Delta_t(\gamma_f+\gamma_\epsilon)},
\end{align}
\end{subequations}
As we can see from above equations, when we set the detuning of the trapping beam to be $\Delta_t=\omega_s$ and set $\gamma_\epsilon/\gamma_f\ll1$, the intracavity field amplitude is strong: $\bar{a}=\bar{b}=\sqrt{2/\gamma_f}\bar{a}_{\text{in}}$ with $\bar{a}_{\text{in}}=\sqrt{P_{\text{trap}}/\hbar\omega'_0}$. The fluctuating field consists of mechanical modulation and quantum fluctuations as:
\begin{widetext}
\begin{subequations}\label{flucfield}
\begin{align}
\hat{a}(\omega)=\frac{-G_b\omega_s\hat{x}+i\omega_s\sqrt{2\gamma_\epsilon}\hat{b}_{\text{in}}+[i(\omega+\Delta_t)-\gamma_\epsilon][-iG_a\hat{x}+\sqrt{2\gamma_f}\hat{a}_{\text{in}}]}
{(\omega+\Delta_t)^2-\omega_s^2-\gamma_\epsilon\gamma_f+i(\omega+\Delta_t)(\gamma_f+\gamma_\epsilon)}\\
\hat{b}(\omega)=\frac{G_a\omega_s\hat{x}+i\sqrt{2\gamma_f}\omega_s\hat{a}_{\text{in}}+[i(\omega+\Delta_t)-\gamma_f][iG_b\hat{x}+\sqrt{2\gamma_\epsilon}\hat{b}_{\text{in}}]}
{(\omega+\Delta_t)^2-\omega_s^2-\gamma_\epsilon\gamma_f+i(\omega+\Delta_t)(\gamma_f+\gamma_\epsilon)}\,,
\end{align}
\end{subequations}
\end{widetext}
with $G_a\equiv G_0\bar{a}$ and $G_b\equiv G_0\bar{b}$ (notice that in case of $\Delta_t=\omega_s$, we have $G_a=G_b$).
The radiation pressure force acting on the trapped mirror is given by
\begin{equation}\label{force}
\hat{F}_{\text{rad}}(\omega)=\hbar [G_a^*\hat{a}(\omega)+G_a\hat{a}^\dagger(-\omega)-G_b^*\hat{b}(\omega)-G_b\hat{b}^\dagger(-\omega)],
\end{equation}
which can be split into two parts:
\begin{equation}
\hat{F}_{\text{rad}}(\omega)=-K_{\text{opt}}(\omega)\hat{x}(\omega)+\hat{F}_{\text{BA}}(\omega)
\end{equation}
The first and second term represent the pondermotive modification of the mechanical dynamics and the back-action quantum radiation
pressure noise respectively. The $K_{\text{opt}}(\omega)$ here is the optomechanical rigidity which can be expanded in terms of $\omega$ if the typical frequency of mechanical motion is smaller than the other frequency scale in the trapping system:
\begin{equation}\label{rig}
\begin{split}
K_{\text{opt}}(\omega)&\approx K_{\text{opt}}(0)+\frac{\partial K_{\text{opt}}}{\partial \omega}\omega+\frac{1}{2}\frac{\partial^2 K_{\text{opt}}}{\partial \omega^2}\omega^2\\
&\equiv m\omega_{\text{opt}}^2-im\Gamma\omega-m_{\text{opt}}\omega^2.
\end{split}
\end{equation}
The first term in \eqref{rig} gives the trapping frequency and the second and third terms give the velocity and acceleration response of the trapped mirror which are optical (anti-)damping $\Gamma$ and
optomechanical inertia $m_{\text{opt}}$, respectively.

Substituting ~\eqref{classfield} and ~\eqref{flucfield} into
 ~\eqref{force} and taking the expansion with respect to detection frequency $\omega$, we can get
 analytical expressions of the optical rigidity and radiation pressure noise. However, they are too cumbersome to show.
In the following, we show approximate results in the interesting parameter region of $\Delta_t\sim\omega_s$ and $\gamma_\epsilon\ll\gamma_f$ in which the back-action noise can be coherently canceled.

\subsection{Dynamics and back-action}
The optical spring frequency is given by:
\begin{equation}\label{spring}
\omega_{\text{opt}}^2=\frac{\hbar G_a^2}{m\omega_s}+\cal{O}(\eta).
\end{equation}
Substitute $\bar{a}_{\text{in}}$, $G_a$, $\omega_s$ and $\gamma_f$ in, and we have Eq.(16) in the main text.
 The $\cal{O}(\eta)$ here describes all the high order terms with $\eta\sim(\Delta_t-\omega_s)/\omega_s,\gamma_\epsilon/\gamma_f$.
Notice that this optical spring can be treated effectively as
a quadratic trap of the vibrating mirror on the anti-node of our trapping beam as shown in \footnote{H. Miao, S. Danilishin, T. Corbitt, and Y. Chen, Phys. Rev. Lett. {\bf 103}, 100402 (2009)}

The optical (anti)-damping factor $\Gamma$ is given by (to $1st$ order of $\omega$):
\begin{equation}\label{damping}
\Gamma=\frac{16\hbar G_a^2}{m\gamma_f\omega_s}\left(\frac{\Delta_t-\omega_s}{\omega_s}\right)-\frac{8\hbar G_a^2\gamma_f}{m\omega_s^3}\left(\frac{\gamma_\epsilon}{\gamma_f}\right)
+\mathcal{O}(\eta^2)
\end{equation}
It is clear from this formula that in the ideal case when
$\Delta_t=\omega_s$ and $\gamma_\epsilon=0$, the optical damping is completely cancelled. Therefore by carefully choosing the system parameters, we can achieve a small positive damping when the
end mirror is not perfectly reflective.

The main contribution to the optomechanical inertia is at zeroth order of $\epsilon$:
\begin{equation}\label{neginertia}
m_{\text{opt}}=-\frac{\hbar G_a^2}{\omega_s^3}+\mathcal{O}(\eta)
\end{equation}
which is extremely small as we have shown in the main text.

Finally, the back-action radiation pressure force noise spectrum is given by:
\begin{equation}\label{residuerp}
S^{\text{rad}}_{\text{FF}}=\frac{2\hbar^2G_a^2\gamma_f}{\omega_s^2}
\left(\frac{\gamma_\epsilon}{\gamma_f}\right)+\mathcal{O}(\eta^2)
\end{equation}
Notice that the back-action force spectrum is zero when the
end mirror is perfectly reflective ($\gamma_\epsilon=0$).

\begin{figure}[!t]
 \includegraphics[width=0.3\textwidth]{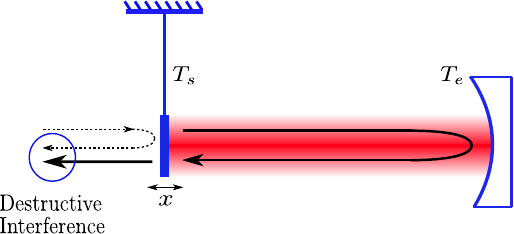}
\caption{Back-action evasion: destructive interference
between the field directly reflected from the oscillating mirror and the field transmitted out of the cavity.
\label{cancel}}
\end{figure}

The physical explanation of this back-action evasion phenomenon is shown in Fig.\ref{cancel}. The part of the outgoing fields which
contains the displacement signal can be written as (\emph{suppose the end mirror
is perfectly reflective}):
\begin{equation}
\hat{a}^m_{\text{out}}=-2iG_a\hat{x}+2i\frac{G_a\omega_s^2}{\Delta_t^2}\hat{x}
\end{equation}
The first term on the right hand-side is the field directly reflected from the trapped mirror while
the second term is the field transmitted out of the cavity. We can see that they cancel when
$\Delta_t=\omega_s$. Therefore in this case the
output field does not contain the $x$-information.

Given the parameters listed in Tab.I of the main text, we use \eqref{spring}-\eqref{residuerp} to calculate the modification of the mechanical dynamics by the trapping beam, and list the effective parameters in  Tab.II of the main text. We can see that velocity response is a mechanical damping factor $\Gamma$ which will not cause instability and is too small to affect the OMIT effective cavity bandwidth $\gamma_{\text{opt}}$. The negative inertia $m_{\text{opt}}$ is also too small to be comparable to the mass of the mechanical oscillator.

\section{Effective temperature of the mirror-endowed oscillator}
Here we estimate the temperature of the mirror-endowed oscillator due to the additional heating caused by optical absorption. We assume the oscillator  to be a silicon cantilever mirror with thickness $h$, Young's modulus $Y$ and density $\rho$. Under the assumption that $l>b,h$, the fundamental frequency of the cantilever is given by \footnote{L. Meirovitch \emph{Analytical Methods in Vibrations.} Prentice Hall (1967)}
\be
\omega_{m0}=1.875^2\sqrt{\frac{YI}{\rho S l^4}},
\ee
where the $S=bh$ and $l$ are the cross-sectional area and length of the cantilever. Then $I=bh^3/12$ is the moment of inertia of the beam cross-section. Using the parameters given in Tab.\,\ref{table:sample A1}, the resonant frequency has the value about $180\text{Hz}$.
\begin{table}[!t]
\caption{Sampling parameters for the trapped oscillator}
\begin{tabular}{ccc}
\hline
$\rho$& mass density &2329 $\text{kg}/\text{m}^3$\\
\hline
$Y$& Young's modulus & $130\text{GPa}$\\
\hline
$l$& cantilever beam length & $1.5\text{mm}$\\
\hline
$h$& cantilever thickness & $0.37\mu$m\\
\hline
$b$& cantilever width &$0.3$mm\\
\hline
\end{tabular}
\label{table:sample A1}
\end{table}

We also assume that the suspended mirror inside the cavity has thermal conductivity $\kappa(T)=\kappa_0T^n$. The cantilever is illuminated by the trapping field with intra-cavity power $P^c_{\text{trap}}$. As a simple 1-D heat transport problem, Fourier's law says that the heat power passing through the cross-section $S=bh$ of the mirror material at distance $z$ from its center equals to $P_{\text{cond}}= -S\kappa T'(z)$. Integrating the heat transport equation from the illuminated spot center with temperature $T_0$ to the boundary with temperature $T$, we have the relation between the $T_0$ and the absorbed power for a rectangular shape mirror  $P_{\text{abs}}$ as:
\be\label{heat transport}
P_{\text{abs}}=\frac{2S\kappa_0}{l(n+1)}(T_0^{n+1}-T^{n+1})
\ee
Typically we have $n\sim2$ at cryogenic temperature$^3$. Using the sample parameters $ 10\text{ppm}$, $P_c\sim15\text{W}$ and the conductivity of the material\footnote{ C. J. Glassbrenner and G. A. Slack, Phy. Rev. 134, A1058 (1964)} $\kappa_0= 10\text{W}/(\text{m}.\text{K}^n)$, we have $T_0= 6\text{K}$ from Eq.\eqref{heat transport}.

How this absorption-induced 6K temperature around the hot spot and its nonuniform distribution across the beam cantilever influences the thermal noise is not entirely clear and needs further study. The loss of the cantilever motion can be classified as surface loss and body loss. The body loss is mainly through the clamping point where the temperature is around the cryogenic environment temperature. The surface loss, on the other hand, influences the cantilever motion through the coupling of the material surface motion with the local thermal bath, which has a raised temperature from the trapping beam heating. Whether the thermal noise due to surface loss degrades the squeezed light or not depends on detailed design of the experiment and needs a more sophisticated study. Moreover, the trapping beam does not illuminate the cantilever beam uniformly thereby a heat flux will be built up across the cantilever beam with temperature gradient about $\nabla T\sim 3\times 10^3 \text{K}/\text{m}$. The non-equilibrium thermal noise associated with this heat flux is also unclear and needs to be addressed in future research.

\end{document}